\documentclass[twocolumn,english,superscriptaddress]{revtex4-1}
\usepackage{epsfig}
\usepackage{color}
\usepackage{tipa}
\usepackage{tipx}
\usepackage{graphicx}
\usepackage{amsmath}
\usepackage{soul}
\makeatletter

\usepackage{babel}
\makeatother

\begin{document}

\title{Searching for the Gardner transition in glassy glycerol}

\author{Samuel Albert}
\affiliation{SPEC, CEA, CNRS, Universit\'e Paris-Saclay, CEA Saclay Bat 772, 91191 Gif-sur-Yvette Cedex, France.}

\author{Giulio Biroli}
\affiliation{Laboratoire de Physique de l'Ecole normale sup\'erieure ENS, Universit\'e PSL, CNRS, Sorbonne Universit\'e, Universit\'e Paris-Diderot, Sorbonne Paris Cit\'e, Paris, France}

\author{Fran\c cois Ladieu}
\affiliation{SPEC, CEA, CNRS, Universit\'e Paris-Saclay, CEA Saclay Bat 772, 91191 Gif-sur-Yvette Cedex, France.}

\author{Roland Tourbot}
\affiliation{SPEC, CEA, CNRS, Universit\'e Paris-Saclay, CEA Saclay Bat 772, 91191 Gif-sur-Yvette Cedex, France.}

\author{Pierfrancesco Urbani}
\affiliation{Universit\'e Paris-Saclay, CNRS, CEA, Institut de Physique Th\'eorique, 91191 Gif-sur-Yvette, France.}

\begin{abstract}
We search for a Gardner transition in glassy glycerol, a standard molecular glass, measuring the third 
harmonics cubic susceptibility $\chi_3^{(3)}$ from slightly below the usual glass transition temperature down to $10K$. According to the mean field picture, if local motion within the glass were becoming highly correlated due to the emergence of a Gardner phase then $\chi_3^{(3)}$, which is analogous to the dynamical spin-glass susceptibility, should increase and diverge at the Gardner transition temperature $T_G$. We find instead that upon cooling $| \chi_3^{(3)} |$ decreases by several orders of magnitude and becomes roughly constant in the regime $100K-10K$. We rationalize our findings by assuming that the low temperature physics is described by localized excitations weakly interacting via a spin-glass dipolar pairwise interaction in a random magnetic field. Our quantitative estimations show that the spin-glass interaction is twenty to fifty times smaller than the local random field contribution, thus rationalizing the absence of the spin-glass Gardner phase. 
This hints at the fact that a Gardner phase may be suppressed in standard molecular glasses, but it also suggests ways to favor its existence in other amorphous solids and by changing the preparation protocol.     
\end{abstract}

\maketitle

At low temperatures, glasses display a set of anomalies compared to their crystalline counterparts. For instance,
the specific heat and thermal conductivity violate the Debye law and 
the vibrational properties are different from the ones predicted by the
Debye theory of phononic excitations \cite{ZP71, MS86}. 
These concomitant phenomena have been investigated extensively both at the theoretical and experimental level 
starting from the 70s' \cite{AHV72, philipps}. The central physical question underpinning this field of research is the nature of the excitations that govern the low temperature physics of amorphous solids. One of the main proposals is that those are associated to disordered independent two-level systems (TLS) \cite{AHV72, philipps}. Although the TLS theory allows to explain many experimental results, some puzzles remain unsolved \cite{LV13}, and theoreticians still wonder on the possible collective nature of the low energy excitations \cite{LV13,lubchenko2007microscopic}. 
The recent solution of simple structural glass models obtained in the limit of infinite spatial dimensions \cite{CKPUZ17,berthier2019gardner} has introduced a new 
 possibility in this research effort: amorphous solids may undergo upon compression or cooling a new kind of phase transition, called Gardner transition, that changes their nature, in particular their low temperature properties. 

Let us first recall the main results of the infinite dimensional solution that are relevant for the problem we focus on. 
Within this approach an amorphous solid is described in terms of a metabasin of configurations in which the liquid remains 
trapped at the glass transition. Since within the mean-field theory (realized in the infinite dimensional limit) barriers between 
metabasins are divergent, amorphous solids correspond to separate ergodic components that can therefore be studied using a thermodynamical formalism \cite{PUZ20}. The main result found in studing infinite dimensional Hard and Harmonic Spheres  \cite{CKPUZ17,BU18} 
 is that these systems undergo a Gardner phase transition when lowering the temperature or increasing the pressure:
 below the critical temperature/above a critical pressure the metabasin associated to the 
 solid formed at the glass transition breaks down in a multitude of glassy states organized in a hierarchical fashion \cite{LB19,DC20, ABP20}. This hierarchy is of the very same nature as that found in the spin glass state in certain mean field spin glass models \cite{MPV87}.
This Gardner phase brings about soft modes \cite{FPUZ15}, diverging susceptibilities and collective excitations \cite{BCJPSZ16, BU16}, and therefore is said to be marginal. Remarkably it plays a central role in the quantitative understanding of the critical properties of three dimensional packings of spheres at jamming \cite{CKPUZ14NatComm}. It is therefore also a possible candidate to explain the anomalous low temperature properties of amorphous solids. 

Whether a Gardner transition takes place for generic model systems is a question that has been investigated in the past few years. Already at the mean field level it has been shown that the emergence of a Gardner phase may depend on the model (interaction potential) and on the cooling procedure; proximity to jamming favors its existence \cite{RUYZ15, BU18} while for some interaction potentials, well annealed glasses do not undergo a Gardner transition upon cooling \cite{SBZ17,SBZ19}. Therefore the emergence of Gardner physics, even at the mean field level, is not generic and may depend on the physical context, interaction potential and preparation details. 
Similar results have also been found in simulations, where  evidences of the Gardner transition have been found mainly in Hard Sphere systems \cite{BCJPSZ16, JY17, SZ18, JUYZ18}. On the experimental side, favorable but somewhat indirect evidences have been reported in granular glasses \cite{SD16}, colloidal glasses \cite{HC20} and in two molecular glasses exhibiting a strong Johari-Goldstein $\beta$ peak \cite{allemands}.  From the theoretical point of view, 
going beyond the realm of mean-field theory and including finite dimensional fluctuations is very challenging: the Gardner transition is alike to the spin glass transition in a field \cite{BU15, SZ18}, for which renormalization group results are not conclusive on the possibility of having a transition in three dimensions \cite{BR80, BU15, CY17, HWGM18}. 

All in all, whether standard molecular glasses display a Gardner phase, or at least some signature of it, remains an open question. The aim of this paper is to address this issue by combining experiments and theory. 
At variance with previous experiments \cite{SD16,HC20,allemands}, we measure the low temperature behavior of the third harmonic susceptibility of glassy glycerol, which is a \emph{direct} smoking gun of the transition and is expected to diverge in correspondence of the Gardner point (see below). 
We do not find any hint of such behavior, therefore excluding the possibility of a transition, at least down to $10K$.
In parallel, from a theoretical point of view, we rationalize our findings using a phenomenological approach: at variance with previous
theoretical approaches that investigated numerically the Gardner transition in 
finite dimension
 \cite{SBZ17,SBZ19}, we build up a phenomenological model of the transition itself and we show that within the assumptions considered in this framework, we cannot expect a Gardner transition in standard fragile molecular glasses, at least in typical experimental conditions.

\begin{figure} 
\includegraphics[keepaspectratio,width=8.3cm]{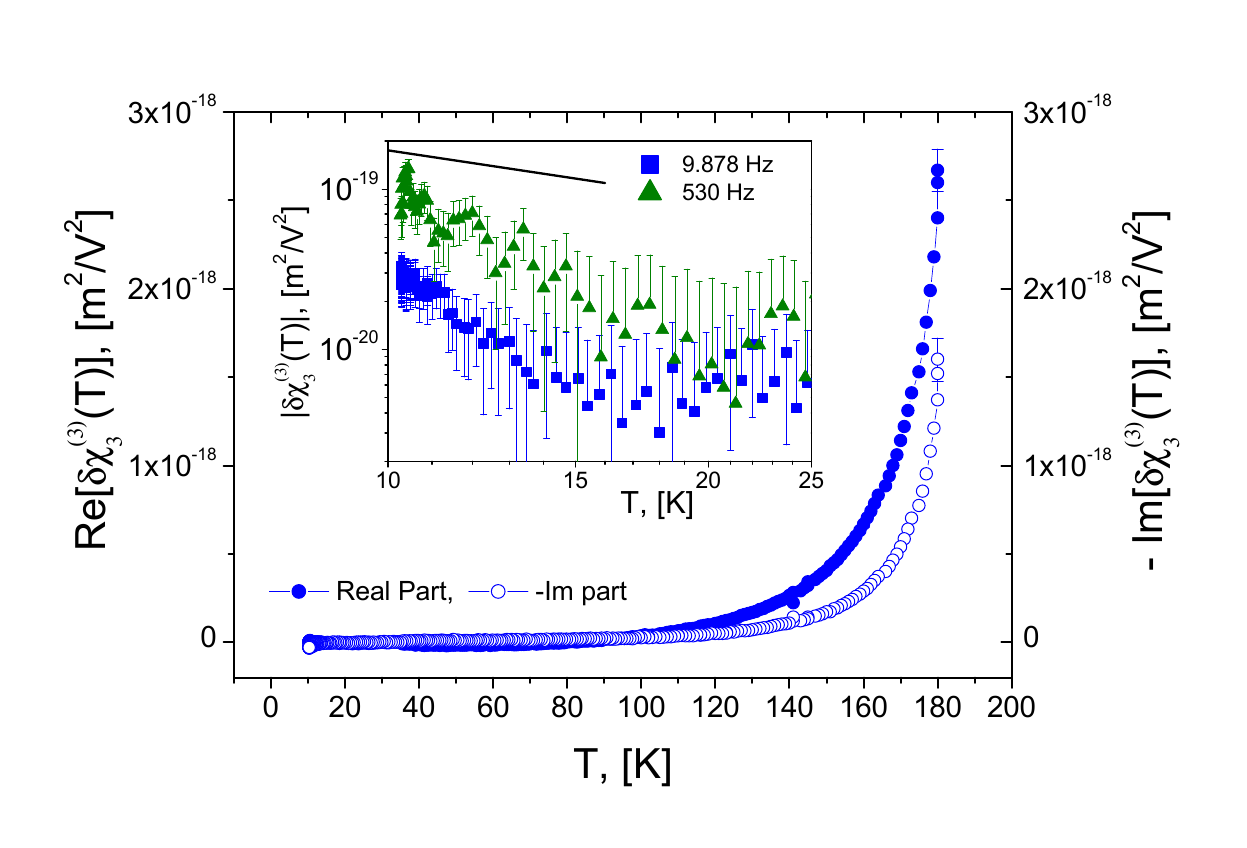}
\caption{(Color Online)  Temperature dependence of $\delta \chi_3^{(3)}(T) \equiv \chi_3^{(3)}(T)-\chi_3^{(3)}(30$K$)$ where $\chi_3^{(3)}$ is the third harmonics cubic susceptibility of glassy glycerol, here measured at an electrical frequency $f=9.878$Hz. The left axis is for the real part data, while the right axis is for the opposite of the Imaginary part data. 
\textit{Inset}: Temperature evolution of $ \vert \delta \chi_3^{(3)}(T,f)\vert$ for $T \le 25$K, where $f=9.878$Hz (squares) or  $f=530$Hz (triangles). The solid line is an example of the $1/T^2$ dependence expected for non interacting TLS's.} 
\label{fig1}
\end{figure}

We start by presenting the results of the experiments on third harmonics cubic susceptibility $\chi_3^{(3)}$ in glassy glycerol from $180K\simeq T_g -8$K, $T_g$ being the usual glass transition temperature, down to $10K$. At low temperature local excitations have a dipolar moment,  $\chi_3^{(3)}$ at fixed angular frequency $\omega $ is expected to 
probe spin-glass order \cite{BY86} and therefore to diverge upon cooling if there is a Gardner transition \footnote{because again, the Gardner transition is the same as a spin glass transition in a magnetic field \cite{BU15}}. Indeed $\chi_3^{(3)}$
 is the dielectric equivalent of the dynamical spin glass susceptibility. 
More precisely, dynamical critical theory leads to \cite{bouchaud2005nonlinear,BBScience16,baity2017matching}: 
\begin{equation}
\chi_3^{(3)}(\omega)=\left(\frac{T_G}{|T-T_G|}\right)^{\nu (2 d_f-d)}g\left(\frac{\omega}{\omega_0}\left(\frac{T_G}{|T-T_G|}\right)^{z\nu} \right)
\end{equation}
where $\omega_0$ is the microscopic frequency, $\nu$ and $z$ are the critical exponents related to the correlation length and to the relaxation time respectively, $g$ a scaling function and $d_f$ the fractal dimension of correlated regions ($d$ is the spatial dimension). Using dynamical scaling, one finds that approaching $T_G$ the third harmonics cubic susceptibility $\chi_3^{(3)}$ should increase when probed at a fixed frequency and it should reach a maximal value of $\chi_3^{(3)}(\omega)\sim (\omega_0/\omega)^{(2d_f-d)/z}$ at $T=T_G$. \\
Henceforth we shall report $\delta \chi_3^{(3)}(T) \equiv \chi_3^{(3)}(T)-\chi_3^{(3)}(30$K$)$. The reason for this substraction is that 
at low temperatures the value of $\vert \chi_3^{(3)} \vert $ is typically $10^4$ times smaller than around the glass transition temperature, i.e. it is so small that the residual spurious third harmonics $V_{source}^{(3)}$ of the voltage source competes with the third harmonics signal of the glycerol sample. Using the fact that the spurious third harmonics does not depend on $T$, we can cancel it out by subtracting the value at the reference temperature $T=30K$. In the Supplemental Material \cite{SM} we present more details and tests that show the efficiency of our experimental procedure.   

In Fig. \ref{fig1} we show the behavior of  $\delta \chi_3^{(3)}$ for a frequency $9.878$Hz as a function of $T$ -note that $\vert \chi_3^{(3)}(30$K$) \vert = (1.0 \pm 0.5) \times 10^{-19}$m$^2/$V$^2$. Our results show a {decrease} from $180K$ to $100K$. Close to the glass transition temperature $T_g$, $\vert \chi_{3}^{(3)} \vert$ probes correlated particle motion \cite{BBScience16,Bru12}.
\begin{figure}
\includegraphics[keepaspectratio,width=8.3cm]{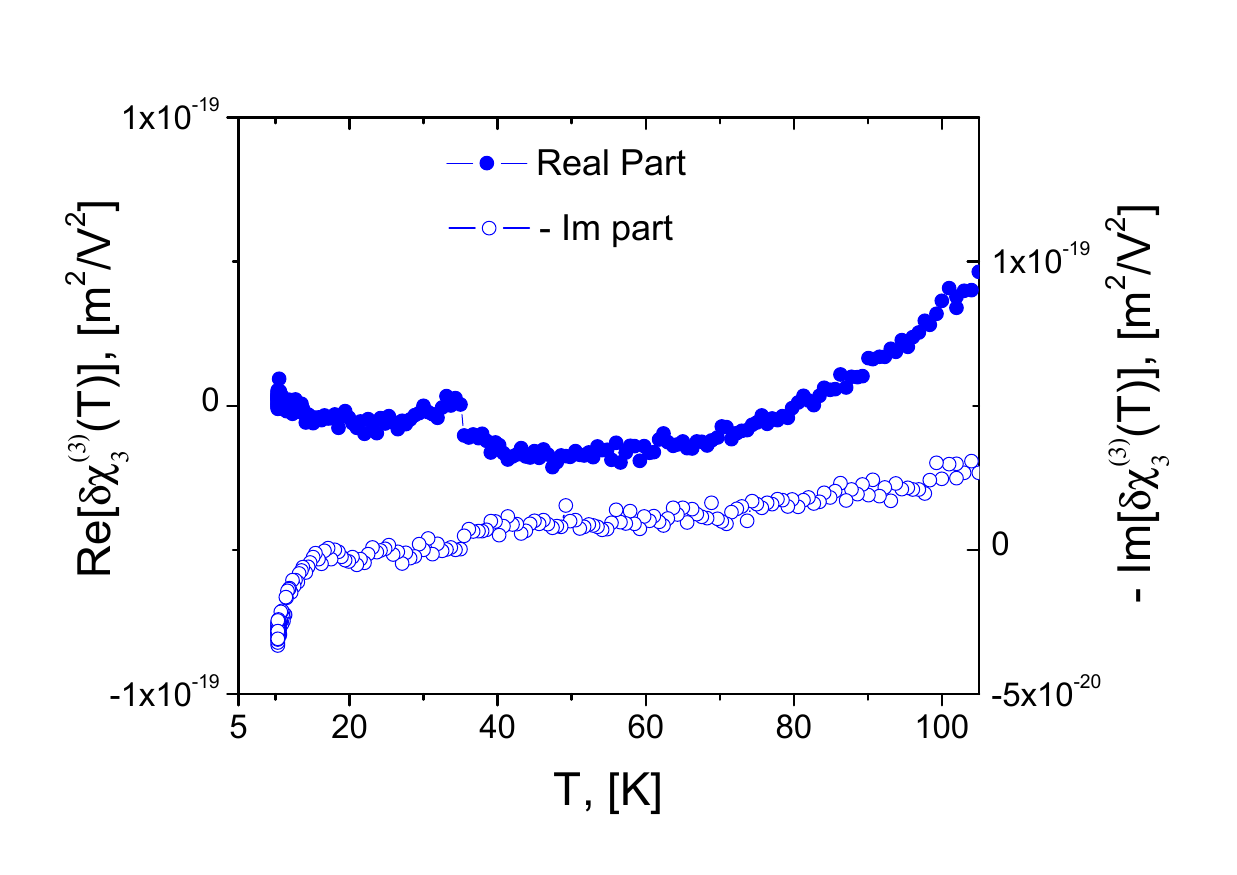}
\caption{(Color Online) Zoom onto the low T part of Fig.1.}
\label{fig2}
\end{figure}

The decrease below $T_g$ is explained as a progressive depletion of mobile regions inside the glass matrix, and does not provide any hint of a Gardner transition. Fig. \ref{fig2} focuses on temperatures below $100K$.
In this regime $|\chi_{3}^{(3)}|$ is essentially constant. A computation of its value based 
on the assumption of {\it independent} local excitations is presented in the Supplemental Material \cite{SM} and leads to a value $0.9 \times 10^{-19}m^2/V^2$ which agrees well with the one found by experiments 
\footnote{We briefly mention that for $T \le 120K$, $|\chi_3^{(3)}|$ is so small that it may be affected by electrostriction (i.e. by  the $E$-induced modulation of the thickness of the sample) and/or by the Kerr effect (i.e. by the $E$-induced variation of the high frequency linear susceptibility). The strongest contribution is that of electrostriction which may lie in the range of a few $10^{-20}$m$^2/$V$^2$ \cite{Bru11,Hansen}. See the Supplemental Material for more details.}.
Note that in the regime $[10$K$; 16$K$]$, one sees a very small increase of $|\delta \chi_3^{(3)}(T,f)|$ upon cooling. This phenomenon, which is hardly above our experimental uncertainty, see the errors bars given in the inset of Fig. \ref{fig1}, was systematically found in the several experiments that we carried out either by varying the value of the electric field $E$ or the angular frequency $\omega$.
It can be explained using TLS theory, which predicts a behavior $\vert \delta \chi_3^{(3)}(T,f) \vert \propto 1/T^2$, see the solid line in the inset of Fig. \ref{fig1} and the Supplemental Material \cite{SM} for more detail. 
All in all, our experimental results do not show any evidence of a Gardner transition from $T_g$ down to $10K$, and they are quantitatively compatible with a scenario based on independent local excitations. Because we are limited to $T\ge 10$K, we cannot strictly exclude that some Gardner transition might happen at a critical temperature below $8$K.

In order to rationalize these findings we use a real space approach. Our main assumption is that 
thermal fluctuations in glasses are due to localized excitations corresponding to partial local atomic motion 
within the frozen glass matrix \footnote{Therefore we start by assuming that TLS are not collective excitations \cite{LV13,lubchenko2007microscopic}.}. This description naturally connects to the one put forward in the past for the low temperature 
properties of molecular glasses, which is based on two level systems (TLS) \cite{AHV72, LV13} as well as to many theories of 
rheology of amorphous solids, which are based on localized soft spots of particles that are prone to rearrangement \cite{YLYWC16,NFMB18}. Besides, recent simulations exhibit   localized excitations in models of molecular glasses \cite{SBZ19}. The interaction between excitations is mediated by the electric and the elastic fields. Since the local conformations 
corresponding to the excitations are random the resulting couplings are random.   
From this real space perspective, the Gardner phase would be a {\it spin-glass phase arising from the interaction of local excitations}.  

In order to study the Gardner phase, we model the localized excitations as $N$ degrees of freedom located in random positions in space. Their density is $\rho=N/V$, where $V$ the total volume of the system.
Each one of them will be denoted $\sigma_{\mathbf x}$, where $\mathbf x$ is the corresponding position. 
Each localized excitation can be in $m_{\mathbf x }$ ($\mathbf x$-dependent) states, which correspond to the possible 
conformations of the localized excitation, i.e of the local atomic positions belonging to the excitation. 
For simplicity, in the following we take $m_{\mathbf x }=2$ for any ${\mathbf x }$ as done for TLS, and use a notation where $\sigma_{\mathbf x}\in\{-1,1\}$
correspond respectively to the low and high energy state of the local excitations. Our arguments and conclusions carry over straightforwardly for $m_{\mathbf x }>2$. The corresponding Hamiltonian reads 
 \begin{equation}
\begin{split}
H&=-U_0 \sum_{i\neq j}\frac{1}{|{\mathbf x}_i-{\mathbf x}_j|^3} u_{ij} \sigma_{{\mathbf x}_i}\sigma_{{\mathbf x}_j}-  \frac 1 2 \sum_{i=1}^N \epsilon_i \sigma_{{\mathbf x}_i}
\label{eq1}
\end{split}
\end{equation}
We have decomposed the interaction between the local excitations in an amplitude, which decreases as the cube of the distance between excitations, and in a random adimensional coupling $u_{ij}$, which depends on the local stress tensors and electric dipoles corresponding to the different states of the local excitations, see e.g. \cite{AHV72, philipps,LV13} for a similar modelling for TLS. $U_0$ is the energy scale of the interaction (measured in temperature times unit of volume). 
The fact that local excitations can be in states with different local energies is encoded in the random positive $\epsilon_i$s. 
The model is effectively a spin glass since the couplings $u_{ij}$ are characterized by an even distribution. This follows from the fact that $u_{ij}$ is bi-linear in the 
dipolar electric moments and the strain tensors associated to the interacting local excitations \cite{AHV72, philipps,LV13}. Since their distribution in space is statistically symmetric under rotation, in particular under a change of sign, the probability of $u_{ij}$ and $-u_{ij}$ are identical.
Note that there are correlations between couplings $u_{ij}$ associated to the same excitations, i.e. $u_{ij}, u_{ik}$
are correlated random variables.  
The local positive energies $\epsilon_i$ are assumed to be independent random variables with a density distribution $\frac 1 \Delta_{typ} f\left(\frac \epsilon \Delta_{typ} \right)$, where $\Delta_{typ}$ is the typical value of $\epsilon_i$ for a localized excitation. We expect, although it is not a crucial ingredient for what follows, that $\Delta_{typ}$ is of the order of the typical effective barrier for $\beta$ relaxation below the glass transition temperature, i.e. thousands of Kelvins. 

Our aim here is not to construct the precise phase diagram of this model, for which a precise characterization of the 
probability distribution of the $u_{ij}$s and $\epsilon_i$s would be required, but instead we want to investigate the possible existence of the Gardner phase based on order of magnitude estimations. In order to do that, one of the key ingredient is the amount of local excitations per unit volume, which can be estimated from TLS physics, since those correspond to very low-energy flank of the distribution. Results on TLS tell us that $f(0)>0$ and that  
 the density of thermally active localized excitations at, say, $10K$ is around $1/(7 \mathrm{nm})^3$ \cite{AHV72, philipps}. Since those 
are expected to be characterized by $\epsilon \ll \Delta_{typ}$, we obtain that 
$$
\int_0^{10K} d\epsilon \frac 1 \Delta_{typ} f\left(\frac \epsilon \Delta_{typ} \right)
\simeq \frac{10K}{\Delta_{typ}} f(0)\simeq \frac{1}{(7 \mathrm{nm})^3}
$$
which sets the scale of $f(x)$ (energies are expressed in units of temperature). Note that assuming a $\Delta_{typ}$ of the order of a few thousands of Kelvin (see above) one gets approximatively an excitation per nanometer cube, which seems reasonable for typical excitations.  
In order to connect with the notation used for TLS, we recall that in that case  $\epsilon = \sqrt{\delta^2+ \Delta_0^2}$ where the potential disorder energy $\delta$ and the coupling tunnel energy $\Delta_0$ are distributed with a density  $\rho(\delta, \Delta_0)=\overline p/\Delta_0$ \cite{AHV72, philipps} where $\overline p$ is a constant. This yields - see Supplemental Material \cite{SM} and \cite{AHV72, philipps}-    $f(0)/\Delta_{typ} \simeq \overline p \ln \left(\Delta_0^{\rm max}/\Delta_0^{\rm min}\right)$ where $\ln\left(\Delta_0^{\rm max}/\Delta_0^{\rm min}\right)\simeq 20$. An important hypothesis for our arguments is that the function $f(x)$ is regular. This amounts to assume that $f(x)/f(0)$ starts from one at $x=0$, varies for $x$ of the order of one and eventually goes to zero for larger $x$s. As far as order of magnitude estimates are concerned, we can use the simple form $f(x)=f(0)\theta(x-1)$, with $\theta(x)$ the Heaviside function.\\
The main issue we wish to address is whether the interaction between local excitations is large enough to lead to a spin-glass phase. In order to work this out, one has to compare the value of the interaction to the local energy difference $\epsilon_i$; if the latter is too strong then a local excitation is subjected to a very strong bias toward the low energy state $\sigma_i=-1$ and its physical behavior is insensitive to the other ones, i.e. no long-range order can be present. 
Note that even though the interaction is a power-law, it is short-ranged as far as spin-glass order is concerned \cite{bray1982mean}, i.e. the effective field due to the interactions with the other local excitations, is dominated by the closest excitations 
\footnote{It is important to make the difference between the existence of a finite temperature phase transition 
toward a spin-glass phase for which the interaction between closest excitations is crucial, and the ground state properties, for which instead we expect that the long range nature of the interaction plays an important role \cite{muller2015marginal}}.\\
Not all local excitations can be considered active. In fact for a given temperature $T$ and a given observation time $\tau$, some of them are frozen out and cannot change state, or they are just too slow to give rise to collective behavior and cannot participate to the putative spin-glass state. In consequence, to be relevant, a local excitation must have  an $\epsilon_i$ less than a 
certain energy $\overline \epsilon(T,\tau)\le \Delta_{typ}$ which depends on $T$ and $\tau$. On general grounds one expects this energy scale  to be less or equal than $\Delta_{typ}$, and to decrease with $T$ and increase with $\tau$ \footnote{
Assuming an Arrhenius law for the flipping time of local excitations, $\tau_f=\tau_0\exp(V/T)$ (with $\tau_0$ is a microscopic time), one finds that imposing 
that for a given temperature the flipping time is less than $\tau$ requires to focus on excitations with barriers less than  $V(T,\tau)=T\log(\tau/\tau_0)$. Since the barrier $V$ is expected to be correlated to $\epsilon_i$ (it has at least to be larger than $\epsilon_i$) the upper bound on $V$ naturally leads to an upper bound on $\epsilon_i$. This dynamical requirement must be combined with a thermodynamical one  stating that $\epsilon_i \leq T$. i.e. that the local excitation must not be frozen -see Supplemental Material \cite{SM} for an explicit calculation in the case of TLS's.}. The precise expression of $\overline \epsilon(T,\tau)$ is not needed for our arguments.

Using the simplified form of $f(x)$ we therefore find that 
the density of active local excitations is $${\mathcal N}(\overline \epsilon(T,\tau))=f(0)\frac{\overline \epsilon(T,\tau)}{\Delta_{typ}}\,.$$
From (\ref{eq1}) the strength of the interaction between the local excitations is ${\mathcal I}=U_0/\ell^3$, where $\ell(T)$ is the typical distance between them. Hence, the interaction strength is proportional to the density of localized excitation, which by the previous equation is proportional to the typical strength $\overline \epsilon(T,\tau)$ of the random fields. These relations therefore allow to establish a direct comparison between $\mathcal I$ and  $\overline \epsilon(T,\tau)$:
\begin{equation}\label{comparison}
\mathcal I(\overline \epsilon(T,\tau))=\frac{U_0}{\ell^3}\simeq U_0{\mathcal N}(\overline \epsilon(T,\tau))
=k\overline \epsilon(T,\tau)
\end{equation}
where $k=U_0f(0)/\Delta_{typ}=U_0\overline p \ln \left(\Delta_0^{\rm max}/\Delta_0^{\rm min}\right)$. 
For molecular glass-former prepared under normal quenched condition $k$ is of the order $0.002-0.02$ \cite{Bur04,Par07}. In the Supplemental Material \cite{SM} we work out this value for glycerol, and show that even considering the additional modes showing up in the Boson peak region, $k$ may reach $0.04$ at most.  
This implies that the strength of the interaction $\mathcal I(\overline \epsilon(T,\tau))$ is generically much smaller than the typical local energy $\overline \epsilon(T,\tau)$ \footnote{We remark that the quantitative estimates rely on the assumptions that TLS are essentially non-interacting.}.  
Therefore we expect that the Gardner spin glass phase should be suppressed 
as we explain now.
Indeed, theoretical studies have shown 
that random fields hamper the existence of long-range order: within the droplet model an infinitesimal random 
field is enough to destroy the spin-glass phase \cite{fisher1988equilibrium}; whereas within mean-field theory a finite field strength, comparable to the coupling strength, is needed \cite{de1978stability}. 
Simulations and experiments have confirmed the negative role of the field: for three dimensional short-range spin-glasses \cite{baity2014dynamical,expFMTO}, if a transition takes place, it does so for field strengths much lower than the coupling strength. For three dimensional dipolar spin-glasses, a model similar to the one studied in this paper, even without a field the existence of long-range spin-glass order is not established \cite{alonso2017nature}, thus making the fate of the spin-glass phase in a field even more uncertain in this case.  All that leads us to the conclusion that in the present case, where 
the interaction strength between local excitations is typically {\it twenty to fifty times} smaller than the value of the local random field, the emergence of the spin-glass phase, and hence of the Gardner phase, is unlikely. 
 
 The natural question that comes out from the conclusions above is why molecular glasses are so different from colloidal and granular ones for which instead strong signatures of Gardner physics have been found 
\cite{BCJPSZ16, SZ18, JY17,JUYZ18,SD16,HC20,allemands}.  
Our results point towards two possible reasons. On the one hand colloids and granular systems are prepared in such a way that the resulting solids are much less annealed, since the time-scale for microscopic motion are much larger ($10^{-6}s$ for colloids and fraction of seconds for granular media). This leads to a much higher density of soft localized excitations, and in consequence to an increase of the interactions term over the random field one, thus favoring the existence of the Gardner phase. On the other hand, the proximity 
to the jamming transition that takes place for both systems also transforms the nature of their excitations. Indeed, at jamming, on top of localized excitations there are also delocalized ones \cite{LDW13,CCPZ15}, which could favor the Gardner transition. How the mechanisms outlined above conspire together to lead to Gardner physics in three dimensional colloidal and granular systems is not clear. Simulations and experiments can help clarify this issue. 
Direct analysis of the nature of excitations, as the ones performed numerically in \cite{scalliet2019nature}, are instrumental. 
Another possibility is studying systems where the two mechanisms above are separated, e.g. ellipsoids or hard spheres 
under SWAP dynamics \cite{BLW18,BIUWZ18}.
To find a Gardner transition in molecular glasses, it would be interesting to find protocols to prepare very poorly annealed systems. Another possibility, is to study network glasses, such as  
 amorphous Silica (${\rm SiO}_2$), whose structure is close to be marginally connected \cite{TDHV00,DLDLW14} and may then display Gardner physics.

\begin{acknowledgments}
	We thank M. Baity-Jesi, L. Berthier, D. R. Reichman, C. Scalliet, F. Zamponi for useful inputs. 
	GB acknowledge support from the Simons Foundation (\#454935, Giulio Biroli).
This  work  was  supported  by  ``Investissements  d'Avenir"  LabEx-PALM  (ANR-10-LABX-0039-PALM).
\end{acknowledgments}

\nocite{Bru11}
\nocite{Hansen}
\nocite{muller2015marginal}

\bibliography{HS-v3}

\begin{center}
\begin{large}
SUPPLEMENTAL MATERIAL
\end{large}
\end{center}

\section{The experimental setup}
\begin{figure} 
\includegraphics[keepaspectratio,width=8.3cm]{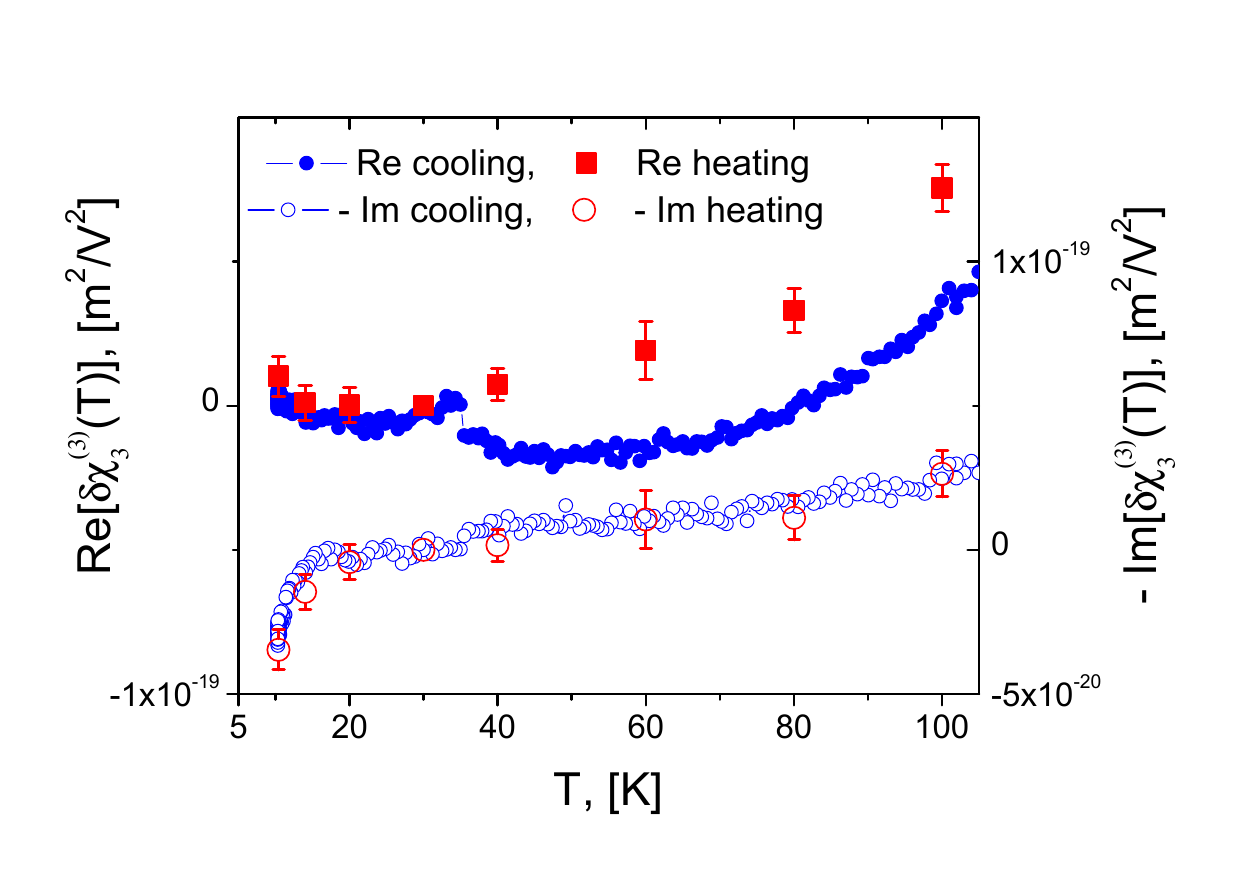}
\caption{(Color Online)  Temperature dependence of the third harmonics cubic susceptibility of glassy glycerol at frequency $\omega=9.878$Hz as a function of temperature. The left axis reports the value of the real part data, while the right axis is for the opposite of the Imaginary part data. On top of the data obtained by the ``cooling'' procedure already reported in Figs.1 and 2 of the main text -full and open circles-, we plot the data recorded in the ``heating'' procedure where the field is systematically varied at each temperature -square symbols-. The two sets of data follow each other, their diffrence is in any case smaller than $3 \times 10^{-20}$m$^2/$V$^2$. We remind the reader that for glycerol around its glass transition temperature $T_g \simeq 190$K, one has $\vert \chi_3^{(3)}(T_g) \vert \simeq 2 \times 10^{-15}$m$^2/$V$^2$.} 
\label{SM-fig1}
\end{figure}

Glycerol was used as received from Sigma Aldrich -purity $99.5\%$- and put in a cell between two stainless steel electrodes separated by a $27.4\mu$m thick Mylar\textregistered  
 ring -see \cite{Thi08,Bru11} for details. The cell was put in a cryocooler allowing to cool the sample down to $10$K. The maximum electrical field $E$ applied to the sample was $13$MV/m. We used the ``twin T'' filter 
method described in Ref. \cite{Thi08}: this suppresses the signal at the fundamental frequency, which is overdominated by the linear susceptibility $\chi_{lin}$, and allows an accurate 
detection of the third harmonics cubic susceptibility $\chi_3^{(3)}$. The experiments reported here were typically carried out as follows: after having been characterized at $205$K$ \simeq T_g+17K$, the sample was cooled, in $6$ hours, down to $10$K -the lowest temperature- at zero electrical field. Then $E$ was varied from $7$MV/m to $13$MV/m -with the frequency fixed to $9.878$Hz- to monitor the third harmonics signal at $T=10$K. The temperature $T$ was then increased to another value -with $E=0$- and after $2$ hours of stabilization, the field $E$ was varied again to monitor $\chi_3^{(3)}$ at the new $T$. The interval from $10$K to $205$K was covered in $13$ successive steps -square symbols in Fig. \ref{SM-fig1}-. In the second part of the experiment -circles in Fig.1 of the main text and in Fig. \ref{SM-fig1}-, the maximum field was applied permanently -still at $f=9.878$Hz- and $T$ was varied continuously from $205$K to $10$K in $6$ hours. 

As explained in the main text, $\chi_3^{(3)}$ becomes so small below -say- $100$K that the $3\omega$ signal coming from the sample is obscured by the one stemming from the  residual spurious third harmonics $V_{source}^{(3)}$ of the voltage source. Because measuring \textit{accurately} $V_{source}^{(3)}$ is difficult, the absolute value of $\vert \chi_3^{(3)} \vert$ is known with an accuracy of $\pm 5\times 10^{-20}$m$^2/$V$^2$, e.g. we find $\vert \chi_3^{(3)}(30$K$) \vert = (1.0 \pm 0.5) \times 10^{-19}$m$^2/$V$^2$. To reduce this uncertainty, we have used the fact that $V_{source}^{(3)}$ does not depend on the temperature, and this is why we have  systematically  plotted $\chi_3^{(3)}(T)-\chi_3^{(3)}(30$K$)$: this difference is known with a better accuracy, as shown by the error bars plotted in Fig. \ref{SM-fig1} and also in the inset of Fig. 1 of the main text. Note that these error bars are well known for the ``heating'' procedure -square symbols of Fig. \ref{SM-fig1}. Indeed in this case, the electrical field is systematically varied at constant $T$, which allows to test that the difference between the measured $3\omega$ signals $V_{meas}(3\omega, T)-V_{meas}(3\omega, 30$K$)$ behaves as expected when varying the field: its phase turns out to be independent of the amplitude $V_{source}^{(1)}$ of the -first- harmonics of the voltage source; while its modulus is found to be proportional to the cube of $V_{source}^{(1)}$. The error bars are deduced from the slight deviations observed with respect to these two requirements about the phase and the cubicity of the modulus.

For the ``cooling'' procedure, reported with circles symbols in Figs.1 and 2 of the main text and in Fig.\ref{SM-fig1}, we cannot perform such a detailed analysis, because the field is kept constantly at its maximum value of $13$MV/m. We observe on Fig. \ref{SM-fig1} that the dispersion between neighbooring points in the cooling procedure is of the same order as the errors bars thoroughly measured in the ``heating'' procedure. Moreover the values obtained in the two procedures nicely correspond for the imaginary part of $\chi_3^{(3)}$ while, for the real part, they show a difference of $\simeq 3 \times 10^{-20}$m$^2/$V$^2$in the $[50$K$;100$K$]$ range. We do not not understand this difference, but we emphasize that it remains very small and cannot not change our conclusions in any respect.

\section{Assessing the value of $\chi_3^{(3)}$ for glycerol at low temperature}

Here we detail how we assess the order of magnitude of the key quantities of our model.

\subsection{Assessing the density of local excitations and their interactions}
In order to evaluate $\mathcal I$ and $\overline \epsilon(T,\tau)$ we use the current available data on TLS estimation in glassy glycerol (that we have used in our experimental setting).

In the standard TLS model, the density of two level systems (TLS) having a potential disorder energy  $\delta $ and a coupling tunnel energy  $\Delta_0$ is distributed with a density  $\rho(\delta, \Delta_0)=\overline p/\Delta_0$. The energy spillting is given by $\epsilon = \sqrt{\delta^2+\Delta_0^2}$. Because we are interested only in ``active'' TLS's, we both take into account: \textit{(i)} a thermodynamic requirement -their gap $\epsilon$ cannot exceed much $k_BT$ otherwise they lie only in their fundamental level-; and \textit{(ii)} a dynamical requirement -their relaxation time must be shorter that the observation time $\tau \simeq 1/f$ where $f$ is the frequency of the electrical field. By integration over $\Delta_0$ one gets the density of TLS $\rho(\epsilon) \simeq \overline p \ln \left(\Delta_0^{\rm max}/\Delta_0^{\rm min}\right)$ where $\Delta_0^{\rm max} \sim k_BT$ -because of requirement \textit{(i)}- and where $\Delta_0^{\rm min}$ is to be calculated by using the dynamical requirement \textit{(ii)}. One shows  \cite{AHV72, philipps} that the relaxation time $\tau_{TLS}$ strongly depends on the values of $\delta, \Delta_0$:  indeed one gets $\tau_{TLS}= \tau_{TLS}(\epsilon, \delta = 0) \epsilon^2/\Delta_0^2$, i.e., for a given gap $\epsilon$, the smaller the tunnel energy $\Delta_0$ the longer the value of $\tau_{TLS}$. In practice $\tau_{TLS}(\epsilon=k_B T, \delta = 0) \sim 1/T^3$ with  $\tau_{TLS}(\epsilon=10K, \delta = 0) \simeq 1$ps. As a result  the logarithmic factor $\ln \left(\Delta_0^{\rm max}/\Delta_0^{\rm min}\right)$ is always in the range $15-25$ and because its variations are often negligible experimentally, one often states that, as we have written in the main paper,  the density of TLS's -per volume and per energy- is $\rho(\epsilon) = \overline p \ln \left(\Delta_0^{\rm max}/\Delta_0^{\rm min}\right)$ with $\ln \left(\Delta_0^{\rm max}/\Delta_0^{\rm min}\right) \simeq 20$. With this convention of inserting the logarithmic factor in the density of states, one gets $\overline \epsilon(T,\tau) = k_B T \ln[\tau/\tau_{TLS}(\epsilon=k_B T, \delta = 0)]/20$ , i.e. $\overline \epsilon(T,\tau)  \simeq k_BT$ up to a factor close to $1$ containing the logarithmic dependence on the observation time $\tau$. 

In the specific case of glycerol, the measurement of specific heat \cite{Ram03} yields $\overline p \simeq 10^{45}$J$^{-1}$m$^{-3}$. Thus, following the arguments presented in the text, the density of TLS active at temperature  $T$ is given by $n_{TLS}(\overline \epsilon(T,\tau))= \overline p \ln\left(\Delta_0^{\rm max}/\Delta_0^{\rm min}\right)\overline \epsilon(T,\tau)$. Therefore the typical distance between them is given by $\ell_{TLS} =\left(n_{TLS}(\overline \epsilon(T,\tau))\right)^{-1/3}$. Therefore the typical interaction strength between TLS is given by $\mathcal I = U_0/{\ell_{TLS}^3}=U_0 \overline p \ln \left(\Delta_0^{\rm max}/\Delta_0^{\rm min}\right)\overline \epsilon(T,\tau)$.
In glassy glycerol it is found that $U_0 \overline p \simeq 10^{-3}$ and therefore one has that $\mathcal I \simeq 2\cdot 10^{-2}\overline \epsilon(T,\tau)$.  Therefore the typical elastic coupling strength between TLS is fifty times smaller than the typical energy scale of local excitations. As discussed in the main text, this argument carries over to localized excitations with energies larger than the usual TLS since the strength of the interaction between them is directly proportional to their local random energy scale, and the proportionality constant is the same one than for TLS. 

Therefore we expect that the Gardner transition can be strongly suppressed by structural disorder.

Additionally we can compute the typical energy $\Delta_{typ}$ corresponding to the extreme case where each molecule of glycerol is a local degree of freedom, i.e. where one imposes ${\mathcal N} =n_{gly}$ with  $n_{gly} \simeq 0.85 \times 10^{28} m^{-3}$ the molecular density of glycerol. One gets -with $k_B$ the Boltzmann constant-:

\begin{equation}
{\Delta_{typ}}/k_B = \frac{n_{gly}}{\left( k_B \overline p \ln\left(\Delta_0^{\rm max}/\Delta_0^{\rm min}\right) \right)} \approx 3\times 10^4 K
\label{eq5}
\end{equation}  

This estimate of $\Delta_{typ}$ is interesting since it turns out to be  larger than (but not so far of) the energy scale $E_{\beta}$ deduced from the activated behavior of the time scale $\tau_{\beta}$ around the glass transition temperature $T_g$: for glycerol one finds indeed $E_{\beta}/k_B \approx 30 T_g \approx 6 \times 10^3 K$.

\subsection{Assessing the density of supplementary excitations in the boson peak region}

On top of the Debye and of the localized excitation contributions, a supplementary contribution shows up in the measured specific heat $\cal C$ in the so called Boson peak region, where one finds a hump in ${\cal C}/T^3$. In glycerol, this hump is visible \cite{Ram03} between $T^\star \simeq 2$K and $10 T^\star \simeq 20$K with a maximum around $8.5$K. Even though the microscopic origin of the Boson peak is still an intense subject of research, we just need here an estimate of the density of excitations involved in it. This is why we use Ref \cite{Par94} where it is argued that the Boson peak comes from the contribution of soft modes, the energy density of which is given by:

\begin {equation}
{\cal D}_{soft} (E) = \frac{\overline p}{6 \sqrt{2}} \left[\frac{E}{1.8 T^\star} \right]^4  
\label{softmodes}
\end{equation}

This yields a contribution ${\cal C}_{soft}$ to the specific heat growing very fast in temperature, namely ${\cal C}_{soft} \sim T^5$. Comparison with experimental data in glycerol shows that this behavior is obeyed up to a $T \simeq 3 T^\star$,  above which some cutoff comes into play, yielding a round maximum in ${\cal C}/T^3$ followed by a decrease at higher temperatures. This is why the sought estimate of the density of supplementary modes is of the order of ${\cal D}_{soft} (3 T^\star)$. Because one finds ${\cal D}_{soft} (3 T^\star)\simeq {\overline p}$ we conclude that the supplementary modes associated to the boson peak region should only double the value of $k$ obtained when considering only the localized excitations, yielding finally a maximal value of $k \simeq 0.04$ as stated in our main text. We emphasize that this is an upper bound for $k$ since we have no logarithmic factor in the range of $20$ involved in the density of soft modes.

\subsection{Assessing the cubic susceptibility of TLS's}
Assuming, on the basis of the previous results, that the localized excitations can be considered as mainly independent objects, the order of magnitude of their contribution to the cubic response is given by:

\begin{eqnarray}
\chi_3^{(3)} &=& \frac{\epsilon_0 \left( \Delta \chi_1\right)^2}{k_B T n(T)} \nonumber \\ 
\ \hbox{\ with\ }\ n(T) &=& k_BT \overline p \ln\left(\Delta_0^{\rm max}/\Delta_0^{\rm min}\right)  
\label{eq6}
\end{eqnarray}

where $\epsilon_0$ is the vacuum dielectric constant, $\Delta \chi_1$ is the contribution of the localized excitations to the static linear dielectric susceptibility, and $n(T)$ is the number of localized excitations per unit volume which are active at temperature $T$. We thus obtain $\chi_3^{(3)} \propto \left(\Delta \chi_1/T \right)^2$. This yields:

$\bullet$ when the standard low temperature behavior for TLS holds $\Delta \chi_1^{TLS} \propto \ln(T/T_{rev})$ with $T_{rev} \approx 0.05K$, $\Delta \chi_1$ hardly varies in temperature and one gets $\chi_3^{(3)} \propto 1/T^2$. Usually the standard low temperature behavior for TLS's is observed up to $\simeq 10$K. Because the re-increase of $\chi_3^{(3)}$ reported in Fig. 2 happens when cooling below $16$K, it may be explained by a contribution of independent TLS's where $\Delta \chi_1$ is fairly constant when $10K \leq T \leq 16K$. 

$\bullet$ when increasing $T$ above $20K$, $\Delta \chi_1$ evolves faster and faster with $T$: in the interval $20K \leq T \leq 80K$ one observes  $\Delta \chi_1 \approx c_1 T$ with $c_1 \simeq 2 \times 10^{-4} K^{-1}$. As a consequence one finds, by using Eq. (\ref{eq6}):

\begin{eqnarray}
\chi_3^{(3)}(20K \leq T \leq 80K) &=& \frac{\epsilon_0 c_1^2}{k_B^2\overline p \ln\left(\Delta_0^{\rm max}/\Delta_0^{\rm min}\right)} \nonumber \\
\  &\approx& 0.9 \times 10^{-19}m^2/V^2
\label{eq7}
\end{eqnarray}
which is idenpendent on $T$ and has the correct order of magnitude with respect to 
the behavior observed in Fig. 1 in the corresponding temperature range.

\subsection{Assessing the contribution of electrostriction and of Kerr effect to the measured cubic susceptibility }
\begin{figure} [h]
\includegraphics[keepaspectratio,width=8.3cm]{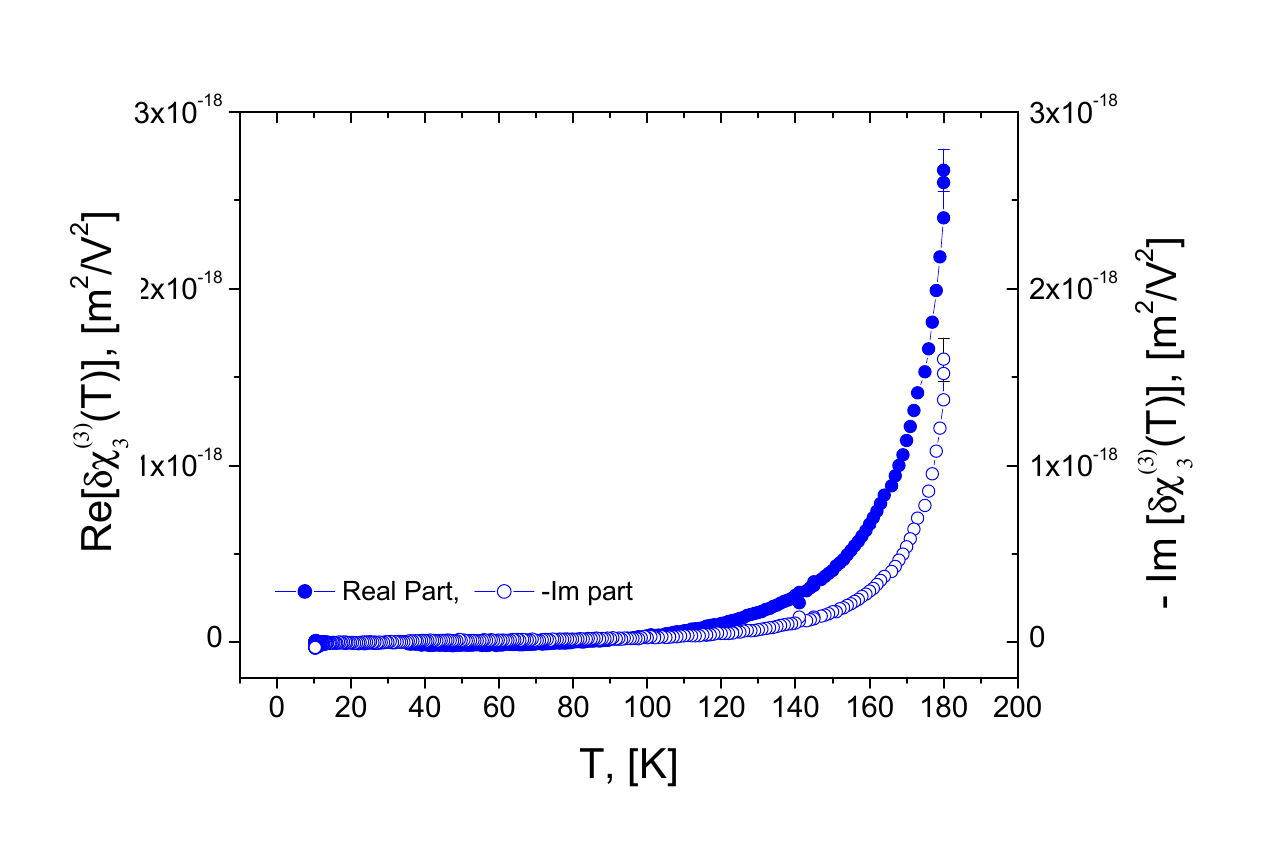}
\caption{(Color Online)  Temperature dependence of the third harmonics cubic susceptibility of glassy glycerol at frequency $9.878$Hz. The left axis is for the real part data, while the right axis is for the opposite of the Imaginary part data.} 
\end{figure}

\begin{figure} [h]
\includegraphics[height= 6.0cm,width=8.3cm]{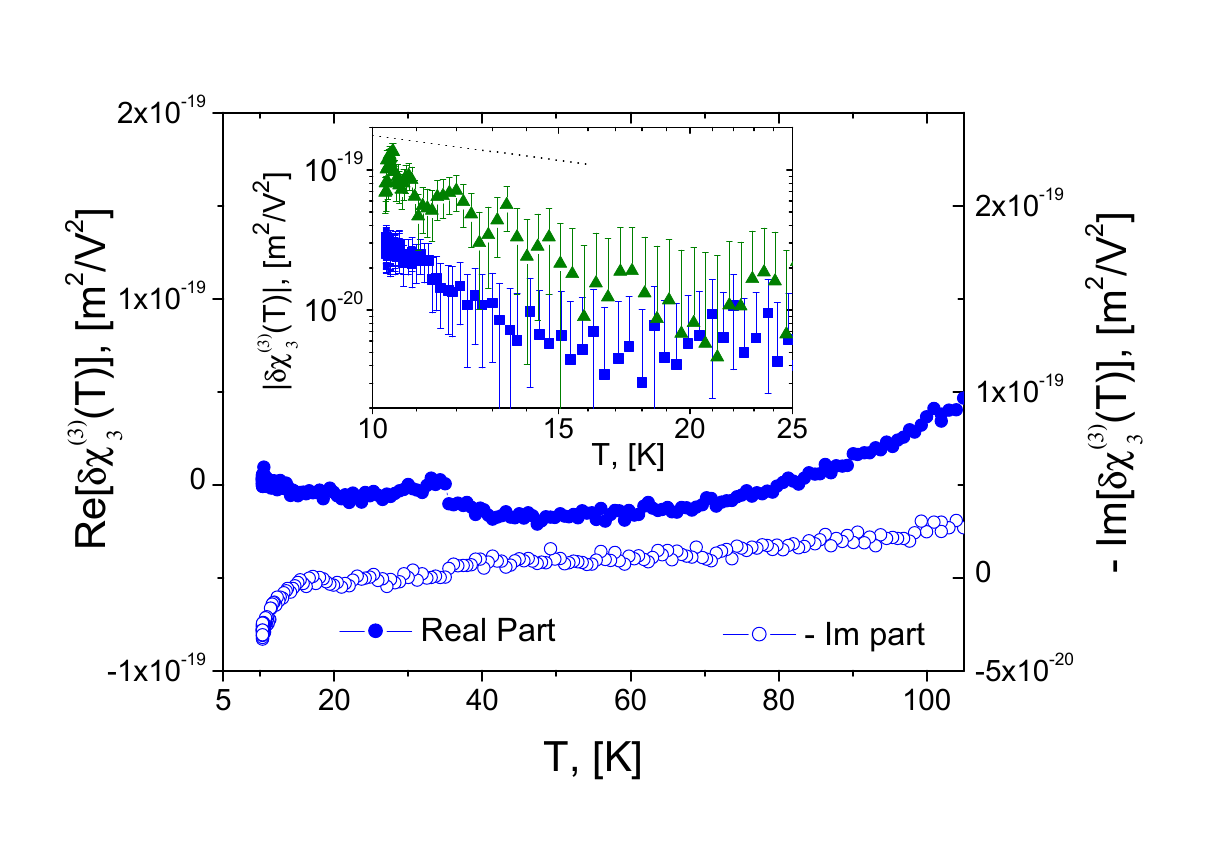}
\caption{(Color Online) Zoom on the $[10K; 100K]$ range (same symbols as in as Fig. 1 of the main text).
\textit{Inset}: Temperature evolution of $ \vert \chi_3^{(3)}(T,f) - \chi_3^{(3)}(30$K$,f)\vert$ for $T \le 25$K. The electrical frequency $f$  is either $9.878$Hz (squares) or  $530$Hz (triangles). The solid line is an example of the $1/T^2$ dependence expected for non interacting TLS's.}
\end{figure}
This order of magnitude of $\chi_3^{(3)} \simeq 10^{-19}m^2/V^2$ 
that we have just derived is so small that it is worth assessing
 the contribution of electrostriction and of Kerr effect 
 which are usually neglected in  the interpretation of low 
 frequency cubic responses of glasses around $T_g$.

Electrostriction comes from the change of the thickness $h$ of the sample arising from the attraction between electrodes due to their opposite charges which create a pressure ${ \Pi}(t) = \epsilon_0 \epsilon_r E^2(t)/2$ with $\epsilon_r =1+ \Delta \chi_1$ the static dielectric constant. As a result the thickness decreases by an amount $\delta h = h_0 {\Pi}/Y$ where $h_0$ is the thickness of the sample at zero applied field and where $Y$ is the effective Young modulus of the sample -i.e. the Young modulus combining that of the glass and that of the spacers separating the electrodes-. Let us write now the polarisation $P$ coming from the linear response $\chi_1$. One has $ P(t)  = \epsilon_0 \chi_{1} V_{source}(t) /(h_0+\delta h)  \simeq \epsilon_0 \chi_{1} V_{source}(t) /h_0 \times (1-\delta h/h_0) $. Inserting $\delta h\propto E^2$, we obtain, on top of the standard linear response, a supplementary term in the dielectric polarisation which is cubic in the field. As a result electrostriction contributes to the cubic susceptibility. More precisely $\delta h$ has two components, a static part $\delta h_0$ and a part $\delta h_{2\omega}$ oscillating at $2 \omega$. Only the latter contributes to the third harmonics susceptibility $\chi_3^{(3)}$ but it is difficult to assess how much dynamical effects damp $\delta h_{2\omega}$ with respect to $\delta h_0$. Therefore we only estimate the effect of $\delta h_0$ and the corresponding contribution of electrostriction $\chi_3^{(1, el)}$ to the first harmonics cubic susceptibility: we obtain $\chi_3^{(1, el)} \approx \frac{\epsilon_0\epsilon_r^2}{3Y}$. This yields $\chi_3^{(1, el)} \approx 0.5 \times 10^{-19}m^2/V^2$ which might be significant with respect to what is reported in Fig.1. However we emphasize that we just have an upper bound here since dynamical damping effects should yield a much smaller electrostriction contribution for the third harmonics susceptibility.

Finaly, we briefly mention Kerr effect, i.e. the fact that the optical index $n_{opt}$ may slightly change upon the 
application of a strong field, yielding a change in the 
high frequency dielectric constant 
$\epsilon_{\infty} = n_{opt}^2$. The strong field may be a 
d.c. field or an optical field. Because the change
 $\delta n_{opt}$ of the optical index is quadratic in the 
 field, the Kerr effect may contribute to the cubic response 
 that is measured in this work. In glasses, most of
  the measurements \cite{Web78} of the Kerr effect have been 
  made by using a strong field of optical origin -by 
  applying typically an intense laser pulse-: using these 
  values to estimate the Kerr contribution 
  $\chi_{3}^{(3,Kerr)}$ to the measured $\chi_{3}^{(3)}$, 
  one finds that 
  $4 \times 10^{-23}m^2/V^2 \leq \chi_{3}^{(3,Kerr)} \leq 4 \times 10^{-21}m^2/V^2$ 
  depending on the considered glassy material. As a result we think that the Kerr contribution can be safely neglected in our experiment.

\end{document}